\documentclass{cimento}

\usepackage{graphicx}  

\title{Higgs Physics as a Proble of New Physics}
\author{S.~Kanemura \from{ins:x}
}
\instlist{
\inst{ins:x}
Department of Physics, University of Toyama, Toyama 930-8555, Japan 
}
\PACSes{\PACSit{12.60.Fr}{\dots}
\PACSit{--.--}{\ldots}}

\begin{document}

\maketitle

\begin{abstract}
The discovery of the Higgs boson at the LHC has opened the door to 
clarify the mechanism of electroweak symmetry breaking and the 
origin of masses of particles.  
The Higgs sector in the SM is the simplest but has no theoretical principle, 
so that there is a possibility of non-minimal Higgs sectors. 
While the standard model is not contradict with the current data 
at the LHC within the error,  most of extended Higgs sectors can also 
reproduce the data.  
An extended Higgs sector often appears in a new physics model beyond the standard model, 
so that we can determine new physics from the Higgs sector. 
In this talk, we discuss various aspects of extended Higgs sectors, 
in particular its phenomenological properties and testability at 
future experiments at the International Linear Collider.
\end{abstract}

\section{Introduction}

Why is the Higgs boson important?  
The Higgs field couples to all the particles in the standard model (SM). 
The Higgs field obtains the vacuum expectation values (VEV) $v$ by 
electroweak symmetry breaking (EWSB), triggered by some unknown dynamics. 
The weak gauge bosons become massive due to the consequence of the 
Higgs mechanism. 
All the quarks and the charged leptons get masses via the Yukawa interactions by 
the replacement of the Higgs field by $v$.  
Even neutrinos (although it is the physics beyond the SM) can have their tiny masses 
through dimension five operators or  neutrino Yulawa couplings after the Higgs 
boson obtains $v$. 
The Higgs field is indeed the origin of mass. 
 
It is also known that the Higgs field is necessary to stabilize the unitarity 
of partial wave amplitudes of elastic scatterings of longitudinally polarized 
weak bosons such as $W_L^+W_L^- \to W_L^+W_L^-$ at high energies. 
Without the Higgs field the S-wave amplitude $a^0(W_L^+W_L^- \to W_L^+W_L^-)$ 
blows up at high energies $a^0 \sim G_F s/(8\pi\sqrt{2})$ where $G_F$ is the Fermi constant and 
$\sqrt{s}$ is the collision energy, and the unitarity is broken at a TeV scale. 
The introduction of the Higgs field cancels such a behavior, and $a^0$ is 
a constant at high energies; $a^0 \sim -G_F m_h^2/(4\pi\sqrt{2})$, where $m_h$ is the mass of the Higgs boson.   
Therefore, the Higgs field is necessary to save the unitarity. 
A condition that the perturbative calculation does not break unitarity gives 
the upper bound such as $m_h < 1$ TeV~\cite{Lee:1977eg}. 

\begin{figure}[t]
\begin{center}
\includegraphics[width=50mm]{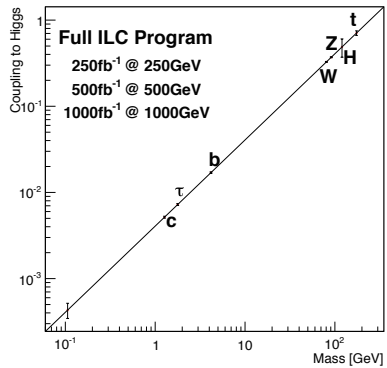}
\end{center}
\caption{
Relation between the mass and the coupling with the Higgs boson in the standard model. 
The expected error precision in the full ILC program is also indicated~\cite{ILC_TDR}.}
\label{fig:c-m}
 \end{figure}

There is no theoretical principle to determine the structure of the Higgs sector within the SM.  
One isospin doublet scalar field $\Phi$ is simply introduced as the minimum form in the SM. 
Under the renormalizability, its potential can be uniquely written as 
\begin{eqnarray}
  V(\Phi) = + \mu^2 |\Phi|^2 + \lambda |\Phi|^4. 
\end{eqnarray} 
By putting an assumption of $\mu^2 < 0$ and $\lambda > 0$, 
the shape of the potential becomes 
like a Mexican hut, and the electroweak symmetry is spontaneously broken at the 
vacuum $\langle \Phi \rangle = (0, v/\sqrt{2})^T$, where  $v \simeq 246$ GeV. 
Consequently, all SM particles but photons and gluons obtain masses from the unique 
VEV $v$. In Fig.~\ref{fig:c-m}, the universal relation between couplings and masses is shown.  

The SM gives a simple description for EWSB. However, the following questions come soon. 
Why is it the miminal form? How we obtain $\mu^2 <0$? What is the origin of the Higgs force $\lambda$?  
Now that a Higgs boson has been found with the mass of about 125 GeV, 
the time has come to consider these questions more seriously. 

\section{Extended Higgs sectors and new physics models}

As there is no principle in the SM Higgs sector, there are many possibilities for 
non-minimal Higgs sectors. 
Notice that while the current LHC data do not contradict the predictions 
in the SM, most of the extended Higgs sectors can also satisfy current data as well. 
These extended Higgs sectors are sometimes introduced 
to give sources to solve the problems beyond the SM such as 
baryogenesis, dark matter and tiny neutrino masses. 
Each scenario does have a specific Higgs sector.  

It is also well known that the introduction of the elementary scalar field is 
problematic, predicting the quadratic divergence in the radiative correction 
to the mass. Such quadratic divergence causes the hierarchy problem. 
There have been many scenarios proposed to solve this problem such as 
Supersymmetry, Dynamical Symmetry Breaking, Extra dimensions and so on. 
Many of the models based on these new paradigms predict specific Higgs sectors 
in their low energy effective theories. 

Therefore, to determine the Higgs sector by experiments is essentially important 
not only to clarify the mechanism of EWSB but also as a window to new physics beyond the SM. 
The discovery of the 125 GeV Higgs boson at the LHC is surely a great step 
for determination of the structure of the Higgs sector. 
 From the detailed study of the Higgs sector, we can determine the model of new physics.  

What kind of extended Higgs sectors we can consider?  
As the SM Higgs sector does not contradict the current data within the errors, 
we may think that there is at least one isospin doublet field. 
An extended Higgs sector can contain additional isospin multiplets 
to the doublet of the SM.  In principle, there can be 
infinite kinds of extended Higgs sectors. 
As a simple example, we may consider models with one additional 
singlet field, one additional doublet field, one additional triplet field and so on.
These extended Higgs sectors can receive constraints from the current data 
of many experiments including those for the electroweak rho parameter and 
for flavor changing neutral currents (FCNCs). 

The rho parameter for a Higgs sector with $N$ multiplets is given at the tree level by  
\begin{eqnarray}
 \rho = \frac{m_W^2}{m_Z^2 \cos^2\theta_W} = \frac{\sum_i \left\{ 4 T_i (T_i+1)- Y_i^2 \right\} |v_i|^2 c_i}
    {\sum_i 2 Y_i^2 |v_i|^2}, 
\end{eqnarray}
where  $T_i$ and $Y_i$  ($i=1, \cdots , N$) are isospin and hyper charges of the 
$i$-th multiplet ($Q_i=T_i+Y_i/2$), and $c_i =1/2$ for real fields ($Y_i=0$) 
and $1$ for completx fields. 
The data shows $\rho=1.0004^{+0.0003}_{-0.0004}$~\cite{PDG}.  
It is found that Higgs sectors with additional doublets $(T_i, Y_i) = (1/2, 1)$  
(and singlets) predict $\rho=1$ at the tree level, like the SM Higgs sector. 
Hence, multi-doublet extensions would be regarded as natural extensions. 
On the other hand, the introduction of higher representation fields 
except for the septet field causes deviations in the rho parameter from unity at the tree level. 
For example, in the model with a triplet field $\Delta$($1,2$) with the VEV $v_\Delta$, 
$\rho \sim 1 - 2(v_\Delta/v)^2$ is given, so that a tuning 
$(v_\Delta/v)^2 \ll 1$ is required to satisfy the data.   
Thus such models are relatively exotic. 
  
It is well known that the multi-Higgs structure receives a severe constraint from the results 
of FCNC experiments. 
FCNC processes such as $K^0 \to \mu^+\mu^-$ and $B^0-\bar{B}^0$ are strongly suppressed~\cite{PDG}.  
In the SM with a doublet Higgs field, 
the suppression of FCNC processes is perfectly explained by the GIM mechanism~\cite{GIM}.
In multi Higgs doublet models where multiple Higgs doublets couple to one quark 
or charged lepton, Higgs boson mediated FCNC can easily occur. 
In order to avoid FCNC, it is required that Higgs bosons have different quantum numbers~\cite{GW}. 

\begin{table}[t]
\begin{center}
\begin{tabular}{|c||c|c|c|c|c|c|}
\hline
 & $\Phi_1$ & $\Phi_2$ & $u_R^i$ & $d_R^i$ & $e_R^i$ &  $Q_L^i$, $L_L^i$   \\
\hline
Type I   &$+$&$-$&$-$&$-$&$-$&$+$  \\
Type II  &$+$&$-$&$-$&$+$&$+$&$+$  \\
Type X &$+$&$-$&$-$&$-$&$+$&$+$\\
Type Y &$+$&$-$&$-$&$+$&$-$&$+$\\
\hline
\end{tabular}
\end{center}
\caption{
Four types of Yukawa interaction in the 2HDM.}
\label{tbl_4type}
\end{table} 

\section{Two Higgs doublet model}

Let us discuss the two Higgs doublet model (2HDM) with $\Phi_1$ and $\Phi_2$, 
the minimal extension with multi-doublet structure.   
For avoiding FCNC, a softly-broken discrete symmetry under 
$\Phi_1\to +\Phi_1$ and  $\Phi_2 \to - \Phi_2$ is imposed~\cite{GW}.  
The Higgs potential is then given by
\begin{eqnarray}
 V &=& + \mu_1^2 |\Phi_1|^2 + \mu_2^2 |\Phi_2|^2 - \mu_{3}^2 (\Phi_1^\dagger \Phi_2 + {\rm h.c.}) \nonumber  \\ 
    && +\lambda_1 |\Phi_1|^4 + \lambda_2 |\Phi_2|^4 +\lambda_3 |\Phi_1|^2|\Phi_2|^2
     +\lambda_4 |\Phi_1^\dagger \Phi_2|^2  
         + \frac{1}{2} \left\{ \lambda_5 (\Phi_1^\dagger \Phi_2)^2 + {\rm h.c.}\right\}. 
\end{eqnarray}
The doublet fields are parameterized as 
\begin{eqnarray}
     \Phi_{i} = \left(\begin{array}{c} 
                                \omega_{i}^{+} \\  
                              \frac{1}{\sqrt{2}}(v_{i} + h_i + i z_i ) \\
                           \end{array}                             
                     \right),         (i=1,2)
\end{eqnarray}
where vacuum expectation values $v_1$ and $v_2$ are expressed by $v$ ($\simeq246$ GeV) 
and $\tan\beta$ by $v^2=v_1^2+v_2^2$ and $\tan\beta=v_2/v_1$. 
The mass matrix of the CP-even scalars is diagonalized by introducing the mixing angle $\alpha$, 
and two mass eigenstates $h$ and $H$ are obtained. The mass matrices of CP-odd and charged 
scalars are diagonalized by $\beta$, and physical mass eigenstates $A$ and $H^\pm$ are obtained, respectively.  
Their masses are given in the decoupling regime ($M \gg v$) by 
\begin{eqnarray}
&&\hspace{-0.6cm}m_h^2 =
\left(\lambda_1 \cos^4\beta +\lambda_2 \sin^4\beta + \frac{1}{2}(\lambda_3+\lambda_4+\lambda_5)\sin^22\beta \right) v^2 + {\mathcal{O}} \left( \frac{v^2}{M^2} \right) ,\nonumber \\
&&\hspace{-0.6cm}m_H^2= M^2+\left(\lambda_1+\lambda_2-2(\lambda_3+\lambda_4+\lambda_5)\right)\sin^2\beta\cos^2\beta \,v^2
+ {\mathcal{O}} \left(\frac{v^2}{M^2}\right) ,\nonumber \\
&&\hspace{-0.6cm}m_{H^\pm}^2= M^2 - \frac{\lambda_4+\lambda_5}{2} v^2,
\hspace{0.6cm}m_A^2= M^2 - \lambda_5 v^2,
\end{eqnarray}
where 
$M$ ($=\sqrt{\mu_3^2/\sin\beta\cos\beta}$) represents 
the soft breaking scale of the discrete symmetry. 

Under the discrete symmetry, there are four possible charge assignments for 
quarks and charged leptons in Table. \ref{tbl_4type}~\cite{Berger}. 
In Type I, all the quarks and charged leptons obtain their masses from $\Phi_1$. 
In Type II, $\Phi_1$ gives masses to down-type quarks and charged leptons, while $\Phi_2$ 
does to the up-type quarks. In Type X, $\Phi_1$ gives mass to the quarks and $\Phi_2$ 
does to charged leptons. 
The rest possibly is called as Type Y.   
The phenomenology for the difference among types of Yukawa interactions have been 
studied in Refs.~~\cite{Aoki:2009ha, Mahmoudi:2009zx}
There are two possibilities to explain the current data which show SM-like. 
When $M^2 \gg v^2$, the additional Higgs bosons are as heavy as $\sqrt{M^2}$, and 
only $h$ stays at the electroweak scale behaving as the SM-like Higgs boson. 
The effective Lagrangian is 
\begin{eqnarray}
 {\mathcal{L}}_{\rm eff} = {\mathcal{L}}_{\rm SM} + {\mathcal O}\left(\frac{v^2}{M^2}\right). 
\end{eqnarray}
Another case is $\sqrt{M^2} \sim v$. In the limit where the $hWW$ coupling takes 
the same value as the SM prediction $\sin(\beta-\alpha)=1$, 
all the Yukawa couplings with $h$ takes the SM values, and $HWW$ is negligible. 
In this case, $h$ behaves as the SM-like Higgs boson. 
When $\sin(\beta-\alpha)$ is slightly smaller than unity, the couplings 
$hVV$ ($V=W$, $Z$), $hff$ ($f=t$,$b$,$c$, $\cdots$)
deviate from the SM predictions depending on type of Yukawa interaction. 
By detecting the pattern of the deviation in each Higgs boson coupling, 
we can distinguish the type of Yukawa coupling in the 2HDMs. 

\section{Fingerprinting of models with future precision data at the ILC}

In 2015, the LHC experiment will restart with the highest energy 14 TeV. 
Extra Higgs bosons in extended Higgs sectors 
can be discovered as long as their masses are 
not too large as compared to the electroweak scale. 
On the other hand, at the International Linear Collider (ILC)~\cite{ILC_TDR}, 
these extended Higgs sectors can also be tested by accurately 
measuring the coupling constants with the discovered Higgs bosons $h$. 
In non-minimal Higgs models, the relation in Fig.~\ref{fig:c-m} does not hold, 
so that we can test the SM by using this relation. 
This is complementary with the direct searches at the LHC.

\begin{figure}[t]
\begin{center}
\includegraphics[width=55mm]{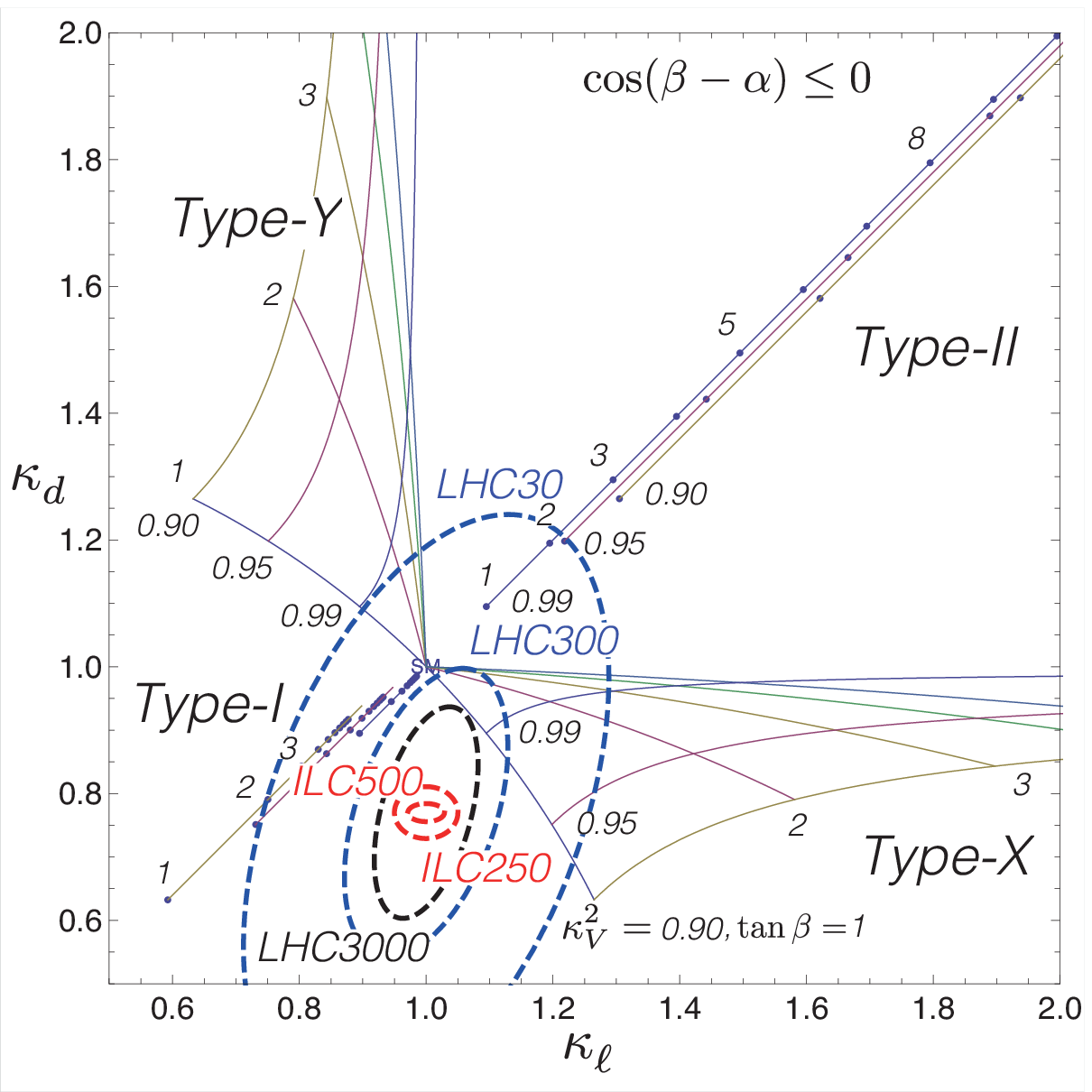}     
\hspace{1cm}
\includegraphics[width=55mm]{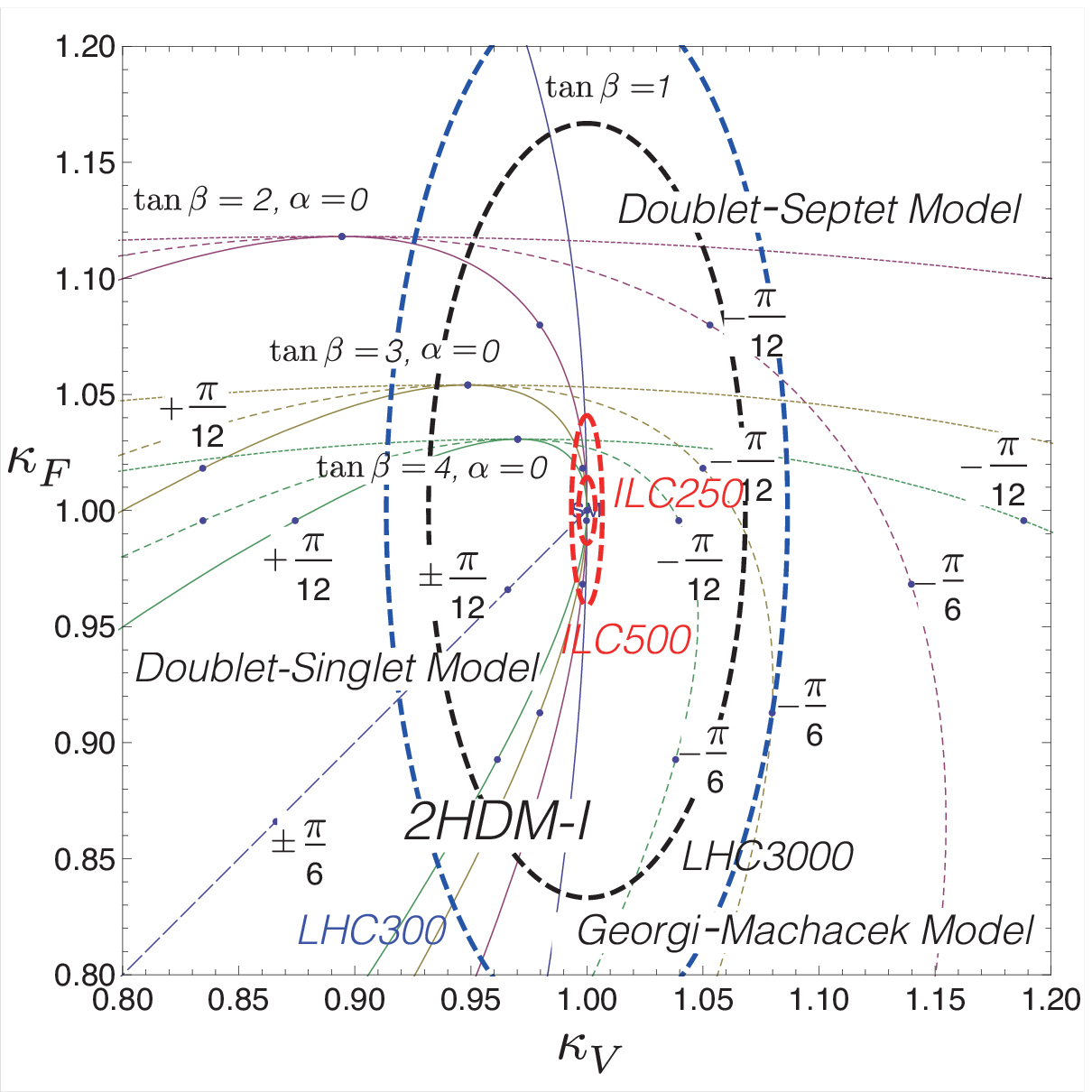}  
\caption{Left: The scaling factors in 2HDM with four types of Yukawa interactions. 
Right: The scaling factors in models with universal Yukawa couplings. The current LHC 
bounds and the expected LHC and ILC sensitivities are also shown at the 68.27 \%  C.L.. 
For details, see the text and Ref.~\cite{Asner:2013psa}}
\end{center}
\label{fingerprint}
\end{figure}

The gauge couplings and Yukawa interactions of $h$ are given by
\begin{eqnarray}
{\mathcal L}^{\rm int}
= +\kappa_W \frac{2m_W^2}{v} hW^{+\mu}W^-_\mu + \kappa_Z \frac{m_Z^2}{v} hZ^\mu Z_\mu 
 -\sum_f\kappa_f\frac{m_f}{v} {\overline f}fh + \cdots , 
\end{eqnarray}
where $\kappa_V$ ($V=W$ and $Z$) and $\kappa_f$ ($f=t,b,c, \cdots$) are the scaling factors measuring 
the deviation from the SM predictions. In the SM, 4 we have $\kappa_V=\kappa_f=1$.

In the 2HDM, $\kappa_V$ are given by
$\kappa_V=\sin(\beta-\alpha)$, while those for the Yukawa interactions are
given depending on the type of Yukawa interaction~\cite{Aoki:2009ha}.
For the SM-like limit $\kappa_V^{}=1$, all the scaling factors  $\kappa_f$ become unity. 
In Fig.~\ref{fingerprint} (Left), the scale factors $\kappa_f$ in the 2HDM 
with the softly-broken symmetry are shown on the  $\kappa_\ell$-$\kappa_d$ plane 
for various values of $\tan\beta$ and $\kappa_V^{}$ ($=\sin(\beta-\alpha)$). 
The points and the dashed curves denote changes of $\tan\beta$ by steps of one.
$\kappa_V$ ($=\kappa_W=\kappa_Z$) is taken as $\kappa_V^2 = 0. 99, 0.95$ and $0.90$.
The current LHC constraints as well as the expected LHC and ILC sensitivities 
for $\kappa_d$ and $\kappa_\ell$  are also shown at the 68.27 \%  Confidence Level (C.L.).
For the current LHC constraints (LHC30), we take the numbers
from the universal fit in Eq.~(18) of Ref.~\cite{Giardino:2013bma}. 
For the future LHC sensitivities (LHC300 and LHC3000), 
the expectation numbers are taken from the Scenario 1 in Table. 1 of Ref.~\cite{CMS:2012zoa}. 
The central values and the correlations are assumed to be the same as in LHC30. 
The ILC sensitivities are taken from Table. 2.6 in Ref.~\cite{ILC_TDR}. 
The same central value without correlation is assumed for the ILC sensitivity curves.
For more details see Refs.~\cite{Asner:2013psa}, and for some revisions see Ref.~\cite{KTYY}.
The analysis including radiative corrections has been done recently~\cite{Kanemura:2014dja}. 

Precision measurements for the couplings of the SM-like Higgs boson $h$ at the ILC 
can also discriminate exotic Higgs sectors. 
In a model with mixing of $h$ with a singlet Higgs field, we have 
a universal suppression on the coupling constants, $\kappa_F^{} = \kappa_V^{} = \cos\theta$ 
with $\theta$ being the mixing angle between the doublet field and the singlet field.
However, $\kappa_F^{} \neq \kappa_V^{}$ is predicted in more complicated 
Higgs sectors such as the 2HDM, the Georgi-Machacek model~\cite{Georgi:1985nv} and 
the doublet-septet model~\cite{Hisano:2013sn}.  
Notice that in exotic models with higher representation
scalar fields such as  the Georgi-Machacek model and doublet-septet model,
$\kappa_V$ can be greater than 1.  
This can be a signature of exotic Higgs sectors.
In Fig.~\ref{fingerprint} (Right), the predictions for the scale factors of the universal 
Yukawa coupling $\kappa_F$ and the gauge coupling $\kappa_V$ are plotted 
in exotic Higgs sectors for each set of mixing angles.
The current LHC bounds, expected LHC and ILC sensitivities 
for $\kappa_F$ and $\kappa_V$ are also shown at the 68.27 \%  C.L..
Therefore, exotic Higgs sectors can be discriminated by measuring $\kappa_V$ and $\kappa_F$ 
precisely. For details, see Ref.~\cite{Asner:2013psa, KTYY}.

\section{Conclusion}

Extended Higgs sectors appear in new physics models beyond the SM. 
We can explore new physics from the structure of the Higgs sector. 
The Higgs sector can be determined by precisely measuring the properties of $h$ 
accurately at the LHC and the ILC. 
In particular, using high ability of the ILC for measuring the Higgs boson couplings,   
we can discriminate extended Higgs sectors,  
and consequently narrow down the new physics models.

\acknowledgments
This talk is partially based on the work with K. Tsumura, H. Yokoya and K. Yagyu~\cite{KTYY}.

 \end{document}